\documentclass{article}

\usepackage{amsmath}
\usepackage{graphicx}

\graphicspath{{./}} 

\newcommand{\Ref}[1]{Ref.~\cite{#1}} 
\newcommand{\Fig}[1]{Fig.~\ref{fig:#1}} 
\newcommand{\Tbl}[1]{Tbl.~\ref{tbl:#1}} 
\newcommand{\Equ}[1]{Eq.~\ref{equ:#1}} 

\renewcommand{\d}{\mathrm{d}} 
\renewcommand{\O}[1]{\mathrm{O}#1} 
\newcommand{\PDF}[1]{\mathrm{PDF}[#1]} 
\newcommand{\Avg}[1]{\mathrm{Avg}[#1]} 
\newcommand{\Var}[1]{\mathrm{Var}[#1]} 
\newcommand{\Skew}[1]{\mathrm{Skew}[#1]} 
\newcommand{\Kurt}[1]{\mathrm{Kurt}[#1]} 

\begin{document}

\title{Statistical distribution of bonding distances in a unidimensional solid}
\author{Roman Belousov \and Paolo De Gregorio \and Lamberto Rondoni \and Livia
Conti}
\date{\today}
\maketitle

\begin{abstract}
We study a Fermi-Pasta-Ulam-like chain with realistic potentials, which models
a unidimensional solid in contact with heat baths at some temperature. We
formulate an explicit analytical expression for the probability density of
bonding distances between neighbor particles, which depends on temperature
similarly to the distribution of velocities. Its validity is verified with a
striking accuracy through simulations.
\end{abstract}


\section{Introduction}

The unidimensional Fermi-Pasta-Ulam-like systems (FPU, \Ref{Gallavotti2008}) are
convenient models for both analytical and computational theoretical studies.
Recently in \Ref{DeGregorio_2011} the open ended FPU-like chain was shown to
mimic closely some thermo-elastic properties of real solids, such as the thermal
expansion and elasticity. Therewith it inspires a certain enthusiasm into
modeling thermodynamic properties of solids by efficient and yet simple means,
e.g. \Ref{Conti_2012}.

We continue to study the one-dimensional chain of point particles, similar to
that of \Ref{DeGregorio_2011}, with one and with both ends opened to allow for
the thermal expansion. We will concentrate the attention mainly on the
statistical distribution of individual phase-space variables. For a particle in
the ideal gas the normal probability distribution of velocity follows from the
Maxwell-Boltzmann statistics, e.g. in \Ref{Sethna2006}. The presence of
interactions between particles through a potential introduces another form of
energy allocation. An interpretation of the same rigor for the Maxwell-Boltzmann
statistics is not available so far in such a general case. However, the Gaussian
distribution of velocities is commonly found and thus consistently assumed to
hold.

The distribution of spatial coordinates in the ideal gas is trivial. Clearly,
the bonding potential energy causes configurational degrees of freedom to take
certain tendencies that give rise to the pair-correlation function. In this
article we investigate our FPU-like chain with Lennard-Jones nearest-neighbor
interactions to find a statistical description of the bonding distances between
particles, which provide for the system of generalized coordinates alternative
to the Cartesian. In fact we show, that it is possible to give an analytical
form for their probability density distribution ($\PDF{r}$), which is connected
to the potential energy ($U_{LJ}$) in a way similar to how the kinetic energy
($K$) enters into the expression for the distribution of velocities ($\dot{x}$):
\begin{equation}
\begin{cases}
   \PDF{r} \propto       & \exp(-\frac{U_{LJ}(r)}{k_B T}) \\
   \PDF{\dot{x}} \propto & \exp(-\frac{K(\dot{x})}{k_B T})
\end{cases}
\end{equation}

\section{Model}
\begin{figure}
\includegraphics[scale=.5]{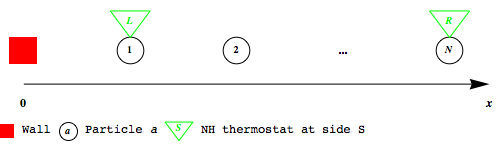}
\caption{Scheme of the open chain model (see text).}
\label{fig:scheme}
\end{figure}

The model consists of $N$ point particles of equal masses $M$ in
one dimension arranged on a horizontal line, as sketched in \Fig{scheme}. The
particles interact solely with their nearest neighbors through the Lennard-Jones
(LJ) potential: $U_{LJ}(r) = E_{LJ} [(r / r_0)^{-12} - 2 (r / r_0)^{-6}]$, where
$r$ is the bonding distance and $E_{LJ}$ is the minimum of potential well at
$r_0$. The particles, indexed in the order of increasing coordinate
$x_a < x_{a+1},\,a=1..N$, obey the following equation of motion:
\begin{equation}
M \ddot{x}_a =
  - \frac{\partial U_{LJ}(x_a - x_{a-1})}{\partial x_a}
  - \frac{\partial U_{LJ}(x_{a+1} - x_a)}{\partial x_a}
\end{equation}
in the chain bulk.

The right-most particle (i.e. the last one from the origin, $a = N$) is in
contact with a deterministic thermostat that operates according to the
Nose-Hoover (NH) scheme at the target kinetic temperature $T$
(\Ref{JeppsRondoni2010}). That is, the equation of motion for the last particle
together with the evolution of the thermostatting variable $\chi_R$ comprise:
\begin{align}
M \ddot{x}_N & = - \frac{\partial U_{LJ}(x_N - x_{N-1})}{\partial x_N}
  - \chi_R M \dot{x}_N \\ \dot{\chi}_R &
  = (K_N - \frac{1}{2} k_B T) / Q
\end{align}
where $K_N = M \dot{x}_N^2 / 2$ is the kinetic energy of the particle and $k_B$
is the Boltzmann constant; $Q$ is an adjustable parameter of thermostat with the
characteristic time $\theta = \sqrt{2 Q / (k_B T)}$.

Analogously, the left-most particle (the first from the origin, $a = 1$) is
coupled with another NH thermostat. However, we explore two settings of the
boundary conditions at the left end. We refer to the \emph{open} chain, when the
first particle interacts by the same Lennard-Jones potential also with a fixed
wall placed at the origin point on the left side. The corresponding equations
are:
\begin{align}
M \ddot{x}_1 &= - \frac{\partial U_{LJ}(x_1)}{\partial x_1}
  - \frac{\partial U_{LJ}(x_2 - x_1)}{\partial x_1}
  - \chi_L M \dot{x}_1 \\ \dot{\chi}_L &
  = (K_1 - \frac{1}{2}k_B T) / Q
\end{align}
We call the chain \emph{free}, when the wall on the left side is removed.
Then the equation of motion for the first particle reduces to:
\begin{equation}
M \ddot{x}_1 = - \frac{\partial U_{LJ}(x_2 - x_1)}{\partial x_1}
  - \chi_L M \dot{x}_1
\end{equation}

Differently from \Ref{DeGregorio_2011}, there are 2 heat baths in our model,
each acting on a single particle. Nonetheless we verified that our results agree
with the global NH thermostat. In our computer experiments the case of 2 heat
baths converged more quickly to the desired steady state. Furthermore, such
setup is rather interesting, for it permits to expose the chain to temperature
differences between the thermostats. Therefore to avoid repeating the same
statements for both cases, our simulations with the global NH thermostat are not
regarded onwards.

The molecular dynamics simulations of the model were performed using the
classical Runge-Kutta integrator of 4th order with a time step $\Delta t$ (e.g.
\Ref{Hairer_2008}). The programming code ensured that the order of particles
relative position was preserved, i.e. $x_a < x_{a+1}$, terminating the execution
otherwise. The basic units of measure, internally adopted in the computer
experiments, were the mass $M$, the length $r_0$, and the unit of simulation
time $\tau$. The derived unit of energy employed further in the text is defined
as $\epsilon = M r_0^2 / \tau^2$. The cartesian coordinates and velocities of
particles were sampled each time after $10^5$ steps of integration to render
$10^4$ snapshots of the phase space.

\Tbl{num_pars} introduces numerical parameters common to all the simulations.
Note the dimensionless definition of Boltzmann constant implying that the
temperature is reported in the units of energy and coincides numerically with
the target average kinetic energy of particles.

\section{Results}\label{sec:rslts}

\begin{figure}
\includegraphics[scale=.5]{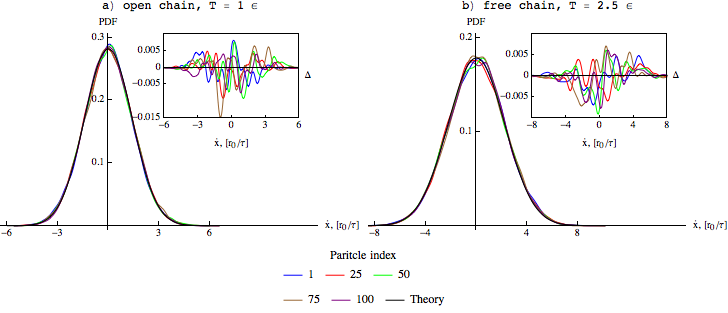}
\caption{Velocity distributions: kernel density estimations for selected
particles from simulations vs. the theoretical curve; the insets depict the
deviation $\Delta = \mathrm{Hist}_a - \PDF{\dot{x}}$ of a smooth histogram
$\mathrm{Hist}_a$ from the theory.}
\label{fig:vel}
\end{figure}

Assumed the system represents the canonical ensemble, the velocity $\dot{x}_a$
of each $a$-th particle is expected to be distributed normally over a
sufficiently long time of simulation. Some smooth histograms constructed from
our simulation data by the kernel density estimation method
(\Ref{Silverman2000}) are adduced in \Fig{vel}: the theoretical curve
corresponds to the Probability Density Function (PDF) $\PDF{\dot{x}} \propto
\exp(-\frac{K_a}{k_B T})$. The accurate coincidence of the predicted curve with
the computational experiment confirms that the chain is in the state of desired
properties.

\begin{figure}
\includegraphics[scale=.5]{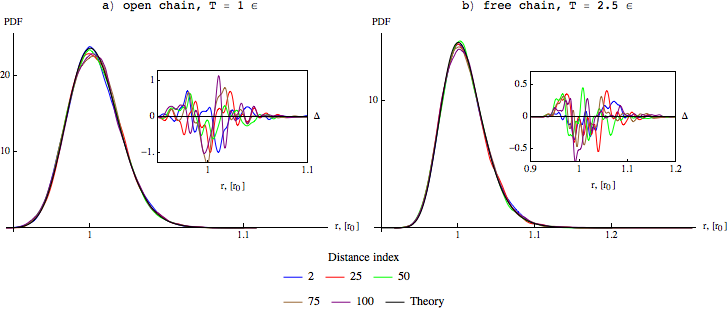}
\caption{Distance distributions: kernel density estimations for selected
bonding distances from simulations vs. the theoretical curve; the insets depict
the deviation $\Delta = \mathrm{Hist}_a - \PDF{r}$ of a smooth histogram
$\mathrm{Hist}_a$ from the theory.}
\label{fig:dst}
\end{figure}

As anticipated in the introduction, the distances between particles
$r_a = x_a - x_{a-1},\,a = 2..N$ (in the case of the open chain $r_1 = x_1$) can
be considered as the configurational degrees of freedom. Their statistical
distribution over snapshots are illustrated in \Fig{dst} by the smooth
histograms alongside the theoretical curve, which is the principle result
of the present article and is discussed in the following.

To construct the analytical expression for the distribution of bonding
distances, we notice that the smooth histograms are bell-shaped curve centered
around $r_0$ with the positive skewness. Thus as the first step we pose
generically a Gaussian function
$\PDF{r} \propto \exp(-\frac{W(r)}{2 \sigma^2})$. As long as $W(r)$ is a
quadratic polynomial of $r$, the very same expression describes the Normal
distribution. The skewness accounts essentially for the thermal expansivity,
causing the average value to shift from the peak of distribution. It can be
controlled by an asymmetric form of $W(r)$.

The asymmetry of interaction potential plays a major role in the thermal
expansion. E.g. the pure harmonic interaction, that corresponds to a quadratic
polynomial, fails to describe realistically the phenomenon. Perhaps a
straightforward way to reproduce this feature through the skewness of expression
being derived is to put $W(r)=U_{LJ}(r)$. Next, the denominator under the
exponent should have units of energy to respect the dimensionless nature of
fraction. By the analogy with the distribution of velocities we take
$2 \sigma^2 = k_B T$ to obtain:
\begin{equation}
\label{equ:family}
\PDF{r} \propto p(r) = \exp(-\frac{U_{LJ}(r)}{k_B T})
\end{equation}

Finally, the expression should be normalized to find the coefficient of
proportionality $c_r$, say. However the integral of the formula built so far
doesn't converge on the support $r \in (0, \infty)$. This difficulty can be
treated in various ways, some of which are discussed in the next section. Here
to circumvent the problem we introduce a cutoff of the support
$r \in (0, 2 r_0]$. It is to note that no cutoff distance was imposed on the
interactions in the simulation model.

Indeed one can see from \Fig{dst}, that PDF drops down to zero very rapidly
and practically vanishes outside the range $0.9 r_0 \leq r \leq 1.1 r_0$.
Therefore the suggested support is more than sufficiently representative and may
be chosen even narrower. After the normalization on the interval
$0 \leq r \leq 2 r_0$, the final formula is obtained:
\begin{equation}
\label{equ:dstr}
\PDF{r} = \begin{cases}
  c_r \exp(-\frac{U_{LJ}(r)}{k_B T}), & \text{if } 0 \leq r \leq 2 r_0 \\
                                   0, & \text{if } r > 2 r_0
\end{cases}
\end{equation}

\begin{figure}
\includegraphics[scale=.5]{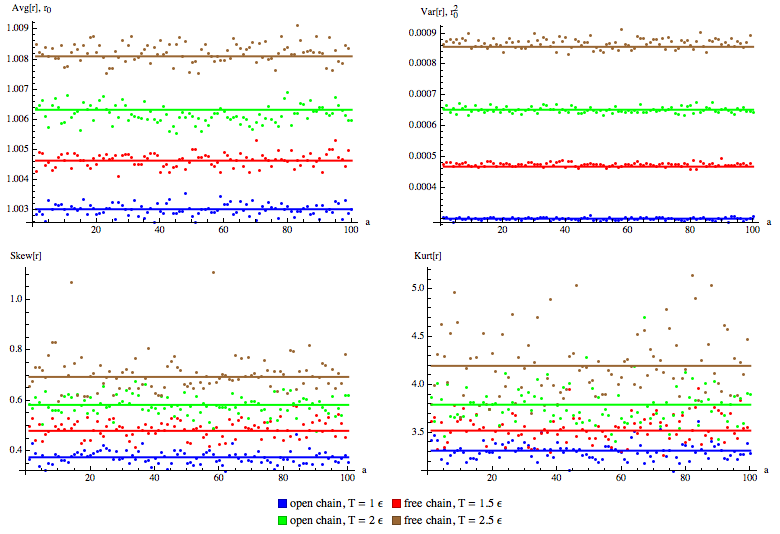}
\caption{Statistics on inter-particle distances along the chain: theoretical
prediction (solid lines) and the estimations from selected simulations (points);
$a$ is the index of distance.}
\label{fig:dst_stats}
\end{figure}

Upon the numerical evaluation of integrals required to normalize \Equ{dstr} at
the proper temperatures, the theoretical curves were built in \Fig{dst}. The
expression turns out to describe very accurately at least first 4 moments of the
distribution, namely: the average value $\Avg{r}$, the variance $\Var{r}$, the
skewness $\Skew{r}$ and the kurtosis $\Kurt{r}$. \Fig{dst_stats} compares the
theoretical moments with the statistics from the simulations for each particle
along. Moreover, they actually appeared almost insensitive to the chosen support
on a broad range of the cutoff values. To enforce the argument, we adduce the
numeric results in \Tbl{dst_stats}, where the discrepancy is seen within the
second or third significant digit between the theoretical values and the
simulation statistics. For excessive lengths of the support notable deviations
begin to emerge from the higher orders to the lowers. This effect is more
evident at elevated temperatures, as shown in \Tbl{dst_stats}.

\section{Discussions}
At first we discuss the alternative procedures to tackle the normalization of
$p(r)$. The problem arises due to the limit
$\underset{r \to \infty}{\lim} p(r) = 1$, which makes the integral
$\int_0^\infty\d r p(r)$ to diverge. A quite efficient solution we elaborated is
to expand $U_{LJ}(r)$ in Taylor series around its minimum at $r_0$ up to an even
power:
\begin{equation}
\label{equ:expansion}
U_{LJ}(r) = c_0 + c_2 (r-r_0)^2 + \dots + c_\alpha (r-r_0)^\alpha
+ \O(r^{\alpha+1}) = U_\alpha(r) + \O(r^{\alpha+1})
\end{equation}
where $\alpha$ is even. By the substitution of the truncated form $U_\alpha$ in
place of $U_{LJ}$, one can recover the normalization on the support
$r \in (0, \infty)$.

\begin{figure}
\includegraphics[scale=.5]{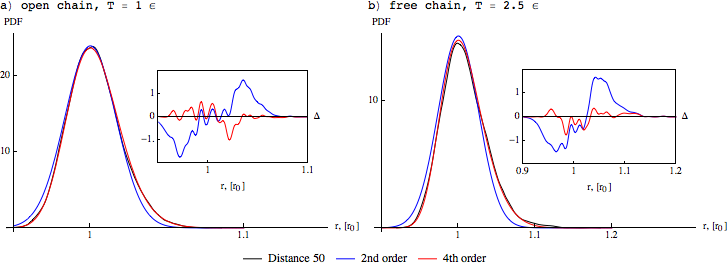}
\caption{Distributions derived from the truncated expansions
of the Lennard-Jones potential in power series around $r_0$; the insets depict
the deviation $\Delta = \mathrm{Hist}_{50} - \mathrm{PDF}_n[r]$ of the smooth
histogram $\mathrm{Hist}_{50}$ from the theoretical curve of order $n$.}
\label{fig:series}
\end{figure}

Obviously, the second order expansion yields a Gaussian, which doesn't describe
appropriately the asymmetry of distribution tails. However, the truncation to
power 4 (or greater) suffices to render a good agreement with the simulations.
\Fig{series} depicts the just stated by comparison of the PDF curves. The curve
of \Equ{dstr} is not reproduced on the plot, because PDF derived from the 4th
order expansion would overlay with it, as well as with the ones procured from
truncations to the higher powers.

Another efficient way of normalization on the unbound interval stemmed from an
idea to form a product $p(r) f(r)$ with a factor $f(r)$ that tends to 0 at
the infinite distance and thus cuts the diverging tail of original distribution.
The function $f(r) = \exp\{- \frac{c_f (r - r_0)^2}{k_B T}\}$ with a constant
$c_f \leq \epsilon / r_0$ having the dimension of energy proved to work fine. We
do not have a more specific prescription to choose the value of constant, as the
sensitivity of results on $c_f$ is really very faint. Although the method
certainly fails for $c_f\gg \epsilon$, it is difficult to propose any reasonably
optimal recommendation, but to set it as small as possible.

The mechanism of the last method can be understood by considering again the
expansion \Equ{expansion}:
\begin{equation}
p(r) f(r) = \exp \{- \frac{c_0 + (c_1 + c_f)(r - r_0)^2 + \O(r^3)}{k_B T}\}
\end{equation}
Thus the factor introduces a correction into the second order term of expansion,
causing the power series to diverge and the exponent to vanish at
$r \to \infty$. The original distribution is recovered for $c_f \to 0$.

By construction, the suggested normalizations are rather convenient
approximations. Both methods produce good numerical predictions on statistics
observed in simulations. Nonetheless \Equ{dstr} perhaps seems more fundamental.
Indeed, one should account that as an artifact of low dimensionality the free
and open chains with Lennard-Jones interactions may get broken sometimes during
the simulation. In such cases, the broken pieces of chain can stand for long
times at distances $r \gg 2 r_0$. Consequently, one would find an outcast
statistics and the distribution \Equ{dstr} would fail. Discarding these cases,
the chain is implicitly assumed to stay bound, i.e. with particles within
certain limits of separation distances.

On the contrary, when the larger distances are admitted in the support, the
normalization constant $c_r$ tends to zero as the integral of $p(r)$ diverges.
In that limit, distances of arbitrary values become equally probable coherently
with the possibility of chain to break. In fact, the bound state we adopt to
model a solid in the equilibrium can be viewed as a metastable state of our
FPU-like chain, which is characterized by the Gaussian distribution of
velocities and \Equ{dstr}.

The bonding distances corresponds to the one-dimensional analogue of the so
called internal coordinates. Once their distribution is known, statistics for
the configurational degrees of freedom in alternative reference frames should be
in principle derivable. E.g., in the open chain the Cartesian coordinate of
$a$-th particle, say, is $x_a = \sum_{b=1}^a r_a$. Then it could be regarded as
a sum of $a$ random variables. It follows though, that the distribution of
Cartesian coordinates would vary by parameters from particle to particle. For
this reason the consideration of distances occasionally is simpler, since they
are directly related to the potential energy. In systems of the higher
dimensionality, the internal coordinates would comprise also the angles.
Therefore the generalization is not so straightforward.

\section{Conclusion}
The presented theoretical developments do not constitute a \textit{de principio}
mathematical derivation for the final expression. Nonetheless the stated
heuristic formulation is strikingly supported by the computational approach with
both numerical and qualitative arguments.

The distribution is valid for the bound state of the FPU-chain with realistic
potentials, which models a solid in equilibrium with heat baths at its
boundaries. The expression degrades consistently to zero, when the chain is
allowed to break and to expand in space arbitrarily.

The statistics of bonding distances is central to many important characteristics
of solids. The distributions themselves are fundamental for account of thermal
vibrations in X-Ray analysis of the atomic structure. Their average values
represent the equilibrium bond lengths. Finally, the distribution of their sum
determines the thermal expansion. The assessment of the analytical form inspires
much interest as the means to analyze and calculate immediately such properties.

\section*{Acknowledgements}
The research leading to these results has received funding from the European
Research Council under the European Community’s Seventh Framework Programme
(FP7/2007-2013) / ERC grant agreement n. 202680.

\bibliographystyle{plain}
\bibliography{ref}

\newpage

\begin{table}
\begin{tabular}{l l}
\hline
Parameter                         & Value               \\
\hline
Number of particles, $N$          & 100                 \\
Depth of potential, $E_{LJ}$      & $ 100 \, \epsilon $ \\
Boltzmann constant, $k_B$         & $2$                 \\ 
Thermostat constant, $Q$          & $10 \, M r_0^2$     \\
Integration time step, $\Delta t$ & $10^{-4} \, \tau$   \\
Total time of simulation          & $10^5 \, \tau$      \\
\hline
\end{tabular}
\caption{Numerical parameters of simulations}
\label{tbl:num_pars}
\end{table}

\begin{table}
\begin{tabular}{l r r r r r r r}
\hline\hline
           & \multicolumn{3}{c}{Distances $r_a$} & \multicolumn{4}{c}{Theory on various supports}                           \\
Statistics & 25 & 50 & 100                       & $(0, 1.2]$ & $(0, 2]$ & $(0,10^2]$ & $(0,10^3]$                          \\ 

\hline\hline
\multicolumn{7}{c}{Open chain, $T = 1\,\epsilon$} \\
\hline
$\Avg{r}\text{, } r_0$                 &  1.00326 &  1.00278 &  1.00303 &  1.00303 &  1.00303 &  1.00303 &  1.00303         \\ 
$\Var{r}\text{, } 10^{-3} \cdot r_0^2$ & 0.298683 & 0.297016 & 0.305084 & 0.299494 & 0.299494 & 0.299494 & 0.299495         \\ 
$\Skew{r}$                             & 0.411562 & 0.349918 & 0.355817 & 0.376281 & 0.376281 &   0.3765 &   2.5735         \\ 
$\Kurt{r}$                             &  3.39459 &  3.24474 &  3.27969 &  3.31398 &  3.31398 &  13.4244 & 1.0 $\cdot 10^6$ \\ 

\hline
\multicolumn{7}{c}{Free chain, $T = 2.5\,\epsilon$} \\
\hline
$\Avg{r}\text{, } r_0$                 &  1.00770 &  1.00756 &  1.00836 &  1.00810 &  1.00811 & 1.02316 & 2.51529          \\ 
$\Var{r}\text{, } 10^{-3} \cdot r_0^2$ & 0.854042 & 0.840263 & 0.893070 & 0.854428 & 0.856988 & 10022.9 & 1.0 $\cdot 10^7$ \\ 
$\Skew{r}$                             & 0.668215 & 0.666151 & 0.784553 & 0.674415 & 0.694886 & 236.632 & 74.7975          \\ 
$\Kurt{r}$                             &  3.98040 &  4.18547 &  4.47063 &  4.00426 &  4.19941 & 59733.9 & 5968.99          \\ 
\hline\hline
\end{tabular}
\caption{Numerical comparison of statistics: selected distances from simulations
and the theoretical predictions based on various cutoffs of the support.}
\label{tbl:dst_stats}
\end{table}

\end{document}